\documentclass[acus]{JAC2003}

\usepackage{graphicx}
\usepackage{booktabs}

\usepackage{dcolumn, mathptmx}
\newcolumntype{.}{D{.}{.}{-1}} % centered on decimal point, including
\newcommand{\be}{\begin{equation}}
\newcommand{\ee}{\end{equation}}
\newcommand{\bea}{\begin{eqnarray}}
\newcommand{\eea}{\end{eqnarray}}

\newcommand{\q}[2]{\ensuremath{#1\ \mathrm{#2}}} % quantity with units
\newcommand{\dt}{\ensuremath{\Delta t}}
\newcommand{\PN}[1]{\ensuremath{P\left(\frac{#1}{w}\right)}}
\newcommand{\e}[1]{\ensuremath{e^{-\left(#1/w\right)^2/2}}}
\begin{document}
\title{MEASUREMENTS OF THE EFFECT OF COLLISIONS ON TRANSVERSE BEAM
  HALO DIFFUSION IN THE TEVATRON AND IN THE LHC}

\author{G.~Stancari,\thanks{E-mail: stancari@fnal.gov.} G.~Annala,
  T.~R.~Johnson, V.~Previtali,
  D.~Still, and A.~Valishev\\
  Fermi National Accelerator Laboratory, Batavia, IL 60510, USA\\
  R.~W.~Assmann,\thanks{Previously at CERN, Geneva, Switzerland.}
  Deutsches Elektronen-Synchrotron, Hamburg,
  Germany\\
  R.~Bruce, F.~Burkart, S.~Redaelli, B.~Salvachua, and G.~Valentino,
  CERN, Geneva, Switzerland}

\maketitle

\begin{abstract}
  Beam-beam forces and collision optics can strongly affect beam
  lifetime, dynamic aperture, and halo formation in particle
  colliders. Extensive analytical and numerical simulations are
  carried out in the design and operational stage of a machine to
  quantify these effects, but experimental data is scarce. The
  technique of small-step collimator scans was applied to the Fermilab
  Tevatron collider and to the CERN Large Hadron Collider to study the
  effect of collisions on transverse beam halo dynamics. We describe
  the technique and present a summary of the first results on the
  dependence of the halo diffusion coefficient on betatron amplitude
  in the Tevatron and in the LHC.
\end{abstract}

\section{INTRODUCTION}

Beam quality and machine performance in circular accelerators depend
on global quantities such as beam lifetimes, emittance growth rates,
dynamic apertures, and collimation efficiencies.  Calculations of
these quantities are routinely performed in the design stage of all
major accelerators, providing the foundation for the choice of
operational machine parameters.

At the microscopic level, the dynamics of particles in an accelerator
can be quite complex. Deviation from linear dynamics can be large,
especially in the beam halo. Lattice resonances and nonlinearities,
coupling, intrabeam and beam-gas scattering, and the beam-beam force
in colliders all contribute to the topology of the particles' phase
space, which in general includes regular areas with resonant islands
and chaotic regions. In addition, various noise sources are present in
a real machine, such as ground motion (resulting in orbit and tune
jitter) and ripple in the radiofrequency and magnet power supplies. As
a result, the macroscopic motion can in some cases acquire a
stochastic character, describable in terms of
diffusion~\cite{Lichtenberg:1992, Chen:PRL:1992,
  Gerasimov:FERMILAB:1992, Zimmermann:PhD:1993, Bruening:PhD:1994,
  Zimmermann:PA:1995, Sen:PRL:1996}.

In studies for the Superconducting Super
Collider~\cite{Irwin:SSC:1989}, the concept of diffusive dynamic
aperture was discussed, and how it is affected by beam-beam forces,
lattice nonlinearities, and tune jitter. Detailed theoretical studies
of beam-beam effects and particle diffusion can be found, for
instance, in Refs.~\cite{Zimmermann:PhD:1993, Bruening:PhD:1994,
  Zimmermann:PA:1995, Papaphilippou:PRSTAB:1999,
  Papaphilippou:PRSTAB:2002}. In Ref.~\cite{Sen:PRL:1996}, the effects
of random fluctuations in tunes, collision offsets, and beam sizes
were studied. Numerical estimates of diffusion in the Tevatron are
given in Refs.~\cite{Sen:PRSTAB:2004, Stern:PRSTAB:2010,
  Previtali:IPAC:2012}.

Two main considerations lead to the hypothesis that macroscopic motion
in a real machine, especially in the halo, may have a stochastic
nature: (1)~the superposition of the multitude of dynamical effects
(some of which stochastic) acting on the beam; (2)~the operational
experience during collimator setup, which generates spikes and dips in
loss rates that often decay in time as $1/\sqrt{t}$, a typically
diffusive behavior.

It was shown that beam halo diffusion can be measured by observing the
time evolution of particle losses during a collimator
scan~\cite{Seidel:1994}. These phenomena were used to estimate the
diffusion rate in the beam halo in the SPS and S$p\bar{p}$S at
CERN~\cite{Burnod:CERN:1990, Meddahi:PhD:1991, Fischer:PRE:1997}, in
HERA at DESY~\cite{Seidel:1994}, and in RHIC at
BNL~\cite{Fliller:PAC:2003}. An extensive experimental campaign was
carried out at the Tevatron in~2011~\cite{Stancari:IPAC:2011.diff} to
characterize the beam dynamics of colliding beams and to study the
effects of the novel hollow electron beam collimator
concept~\cite{Stancari:PRL:2011}. Recently, the technique was also
applied to measure halo diffusion rates in the LHC at
CERN~\cite{Valentino:PRSTAB:2013}. These measurements shed light on
the relationship between halo population and dynamics, emittance
growth, beam lifetime, and collimation efficiency. They are also
important inputs for collimator system design and upgrades, including
new methods such as channeling in bent crystals or hollow electron
lenses.

Halo diffusion rates were measured under various experimental
conditions. In this paper, we focus on the comparison between
colliding and separated beams, in an attempt to expose the effects of
beam-beam forces. After briefly describing the method of small-step
collimator scans, we present data on the dependence of the transverse
beam halo diffusion coefficient on betatron amplitude in the Tevatron
and in the LHC.

\section{EXPERIMENTAL METHOD}

\begin{figure}[b!]
\begin{center}
\includegraphics[width=0.8\columnwidth]{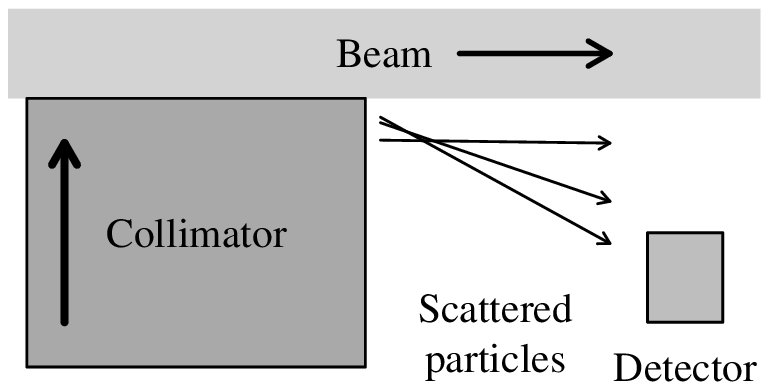}\\
\includegraphics[width=0.8\columnwidth]{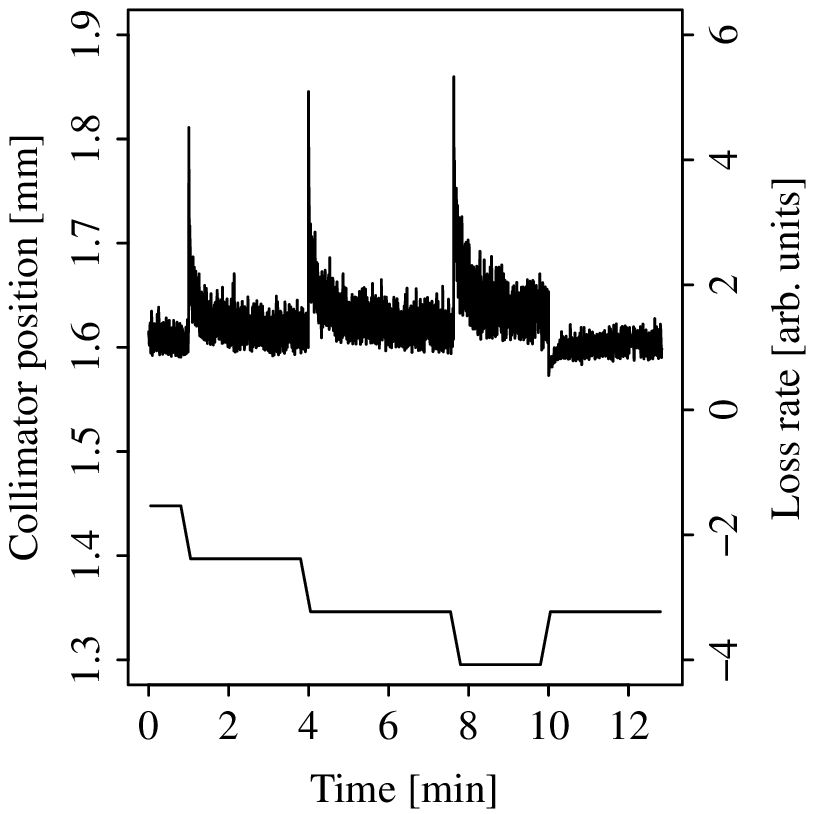}
\caption{Schematic diagram of the apparatus (top). Example of the
  response of local loss rates to inward and outward collimator steps
  (bottom).\label{fig:exp-method}}
\end{center}
\end{figure}

A schematic diagram of the apparatus is shown in
Fig.~\ref{fig:exp-method} (top).  All collimators except one are
retracted.  As the collimator jaw of interest is moved in small steps
(inward or outward), the local shower rates are recorded as a function
of time. Collimator jaws define the machine aperture. If they are
moved towards the beam center in small steps, typical spikes in the
local shower rate are observed, which approach a new steady-state
level with a characteristic relaxation time
(Fig.~\ref{fig:exp-method}, bottom). When collimators are retracted,
on the other hand, a dip in loss rates is observed, which also tends
to a new equilibrium level.  By using the diffusion model presented
below, the time evolution of losses can be related to the diffusion
rate at the collimator position. By independently calibrating the loss
monitors against the number of lost particles, halo populations and
collimation efficiencies can also be estimated.  With this technique,
the diffusion rate can be measured over a wide range of amplitudes. At
large amplitudes, the method is limited by the vanishing beam
population and by the fast diffusion times. The limit at small
amplitudes is given by the level of tolerable loss spikes.

\section{MODEL}

A diffusion model of the time evolution of loss rates caused by a step
in collimator position was developed~\cite{Stancari:diff:2011}. It
builds upon the model of Ref.~\cite{Seidel:1994} and its assumptions:
(1)~constant diffusion rate and (2)~linear halo tails within the range
of the step. These hypotheses allow one to obtain analytical
expressions for the solutions of the diffusion equation and for the
corresponding loss rates vs.\ time. The present model addresses some
of the limitations of the previous model and expands it in the
following ways: (a)~losses before, during, and after the step are
predicted; (b)~different steady-state rates before and after are
explained; (c)~determination of the model parameters (diffusion
coefficient, tail population gradient, detector calibration, and
background rate) is more robust and precise.

Following Ref.~\cite{Seidel:1994}, we consider the evolution in
time~$t$ of a beam of particles with phase-space density~$f(J,t)$
described by the diffusion equation $\partial_t f = \partial_J \left(D
  \, \partial_J f \right)$, where~$J$ is the Hamiltonian action
and~$D$ the diffusion coefficient in action space. The particle flux
at a given location $J=J'$ is $\phi = -D \cdot \left[ \partial_J f
\right]_{J=J'}$.  During a collimator step, the action~$J_c = x^2_c /
(2 \beta_c)$, corresponding to the collimator half gap~$x_c$ at a ring
location where the amplitude function is~$\beta_c$, changes from its
initial value~$J_{ci}$ to its final value~$J_{cf}$ in a time~\dt. The
step in action is $\Delta J \equiv J_{cf} - J_{ci}$. In the Tevatron,
typical steps in half gap were \q{50}{\mu m} in 40~ms; smaller steps
(\q{10}{\mu m} in 5~ms, typically) were possible in the LHC. In both
cases, the amplitude function was of the order of a hundred meters.
It is assumed that the collimator steps are small enough so that the
diffusion coefficient can be treated as a constant in that
region. If~$D$ is constant, the local diffusion equation becomes
$\partial_t f = D \, \partial_{JJ} f$.  With these definitions, the
particle loss rate at the collimator is equal to the flux at that
location:
\be
L = -D \cdot \left[ \partial_J f \right]_{J=Jc}.
\label{eq:flux}
\ee
Particle showers caused by the loss of beam are measured with
scintillator counters or ionization chambers placed close to the
collimator jaw. The observed shower rate is parameterized as
\be
S = kL + B,
\label{eq:obs.rate}
\ee
where~$k$ is a calibration constant including detector acceptance and
efficiency and~$B$ is a background term which includes, for instance,
the effect of residual activation. Under the hypotheses described
above, the diffusion equation can be solved analytically using the
method of Green's functions, subject to the boundary condition of
vanishing density at the collimator and beyond. Details are given in
Ref.~\cite{Stancari:diff:2011}.

Local losses are proportional to the gradient of the distribution
function at the collimator. The gradients differ in the two cases of
inward and outward step, denoted by the~$I$ and~$O$ subscripts,
respectively:
\bea
\lefteqn{\partial_J f_I(J_c, t) = -A_i + 2(A_i - A_c) \PN{-J_c} +
  \mbox{}} \label{eq:gradI} \\
& &
  \frac{2}{\sqrt{2\pi} w} \left\{ -A_i(J_{ci}-J_c) +
 (A_i J_{ci}-A_c J_c) \e{J_c} \right\} \nonumber \\
\lefteqn{\partial_J f_O(J_c, t) = -2 A_i \PN{J_{ci}-J_c}
 + 2 (A_i - A_c) \PN{-J_c} + \mbox{}} \nonumber \\
& & \frac{2}{\sqrt{2\pi} w} (A_i J_{ci} - A_c J_c) \e{J_c}.
\label{eq:gradO}
\eea
The positive parameters~$A_i = -\left[ \partial_J f
\right]_{J=J_{ci}}$ and $A_f = -\left[ \partial_J f
\right]_{J=J_{cf}}$ are the opposite of the slopes of the distribution
function before and after the step, whereas~$A_c$ varies linearly
between~$A_i$ and~$A_f$ as the collimator moves. The parameter~$w$ is
defined as $w \equiv \sqrt{2 D t}$. The function~$P(x)$ is the
S-shaped cumulative Gaussian distribution function: $P(-\infty)=0$,
$P(0)=1/2$, and $P(\infty)=1$.

The above expressions (Eqs.~\ref{eq:gradI} and~\ref{eq:gradO}) are
used to model the measured shower rates. Parameters are estimated from
a fit to the experimental data. The background~$B$ is measured before
and after the scan when the jaws are retracted. The calibration
factor~$k$ is in general a function of collimator position, and can be
determined independently by comparing the local loss rate with the
number of lost particles measured by the beam current transformer.
The fit parameters $(k D A_i)$ and $(k D A_f)$ are the steady-state
loss rate levels before and after the step. The diffusion
coefficient~$D$ depends on the measured relaxation time and on the
value of the peak (or dip) in loss rates.

The model explains the data very well when the diffusion time is long
compared to the duration of the step. The model can be extended by
including a separate drift term (from the Fokker-Planck equation) or a
nonvanishing beam distribution at the collimator.

\section{RESULTS}

\begin{figure}[b!]
\begin{center}
\includegraphics[width=\columnwidth]{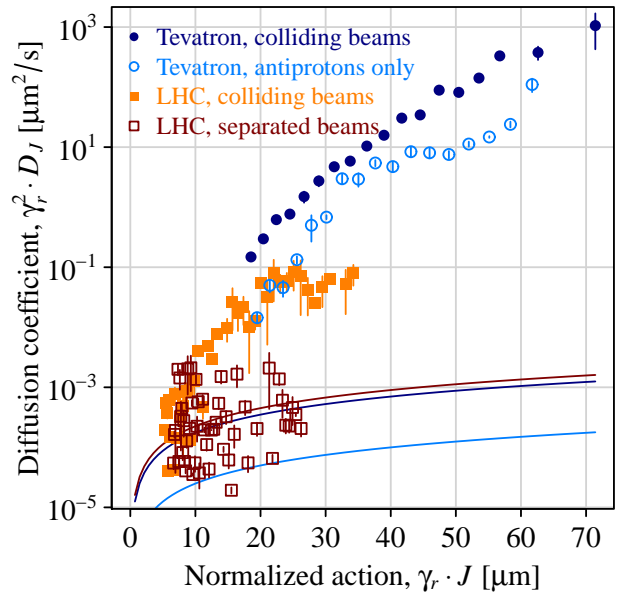}
\caption{Measurements of vertical halo diffusion in the Tevatron and
  in the LHC.\label{fig:diff-meas}}
\end{center}
\end{figure}

All Tevatron scans were done vertically on antiprotons, either at the
end of regular collider stores (0.98~TeV per beam) or with only
antiprotons in the machine at the same top energy. Losses were
measured with scintillator paddles located near the collimators. (A
detailed description of the Tevatron collimation system can be found
in Ref.~\cite{Mokhov:JINST:2011}.)

The LHC measurements were taken in a special machine study at 4~TeV
with only one bunch per beam, first with separated beams and then in
collision, with vertical crossing in the first interaction point (IP1)
and horizontal crossing at IP5~\cite{Valentino:PRSTAB:2013}. Losses
were measured with ionization chambers. Because of the negligible
cross-talk between loss monitors, it was possible to simultaneously
scrape proton beam~1 vertically and proton beam~2 horizontally.

Figure~\ref{fig:diff-meas} shows a comparison of vertical beam halo
diffusion measurements in the Tevatron and in the LHC, for inward
collimator steps. To account for the different kinetic energies of the
two machines, diffusion coefficients are plotted as a function of
normalized vertical collimator action~$I \equiv \gamma_r J$,
where~$\gamma_r$ is the relativistic Lorentz factor. On the vertical
axis, we plot the diffusion coefficient in normalized action space
$D_I \equiv \gamma_r^2 D$, which stems from recasting the diffusion
equation as follows: $\partial_t f = \partial_J \left(D \, \partial_J
  f \right) \to \partial_t f = \partial_I \left(D_I \, \partial_I f
\right)$.

The dark blue filled points refer to the end of Tevatron collider
Store~8733 (13~May~2011). The light blue data (empty circles) was
taken during a special antiproton-only fill (Store~8764,
24~May~2011). The LHC data was taken on 22~June~2012 and refers to
beam~1 (vertical) with separated beams (empty red squares) and in
collision (filled orange squares). The continuous lines represent the
diffusion coefficients derived from the measured core geometrical
emittance growth rates~$\dot{\varepsilon}$: $D = \dot{\varepsilon}
\cdot J$. (In this particular data set, the synchrotron-light
measurements were not sufficient to estimate emittance growth rates of
colliding beams in the LHC.)

In the LHC, separated beams exibited a slow halo diffusion, comparable
with the emittance growth from the core. This fact can be interpreted
as a confirmation of the extremely good quality of the magnetic fields
in the machine. Collisions enhanced halo diffusion in the vertical
plane by about 1~to~2~orders of magnitude. No significant diffusion
enhancement was observed in the horizontal plane. The reason for this
difference is not understood. In the Tevatron, the comparison between
halo and core diffusion rates suggests that single-beam diffusion at
these large amplitudes is dominated by effects other than residual-gas
scattering and intrabeam scattering, pointing towards field
nonlinearities and noise (including tune modulation generated by
power-supply ripple). At the end of the store, collisions enhance
diffusion by about 1~order of magnitude.

From the measured diffusion coefficients, estimates of impact
parameters on the primary collimator jaws are
possible~\cite{Seidel:1994}. One can also calculate the particle
survival time vs.\ amplitude. The diffusion coefficient is related to
the steady-state density of the beam tails, which can therefore be
deduced with a procedure that is complementary to the conventional
static model based on counting the number of lost particles at each
collimator step. These and other consequences of beam halo diffusivity
will be investigated in separate reports.

\section{CONCLUSIONS}

The technique of small-step collimator scans was applied to the
Fermilab Tevatron collider and to CERN's Large Hadron Collider to
study transverse beam halo dynamics in relation to beam-beam effects
and collimation. We presented the first data on the dependence of
transverse beam halo diffusion rates on betatron amplitude. In the
Tevatron, vertical antiproton diffusion at the end of a collider store
was compared with a special store with only antiprotons in the
machine. Even with a reduced beam-beam force, the effect of collisions
was dominant. A comparison with core emittance growth indicated that
halo diffusion of single beams was driven by nonlinearities and noise,
and not by residual-gas or intrabeam scattering. In the LHC,
horizontal and vertical collimator scans were performed during a
special machine study with only one bunch per beam (no long-range
beam-beam interactions). With separated beams, no significant
difference was observed between halo and core diffusion, which
indicated very low noise levels and nonlinearities. In collision,
horizontal diffusion was practically unchanged; the vertical diffusion
rate enhancement was a function of action and reached about~2 orders
of magnitude.  In general, it was confirmed that collimator scans are
a sensitive tool for the study of halo dynamics as a function of
transverse betatron amplitude.

\section{ACKNOWLEDGEMENTS}

The authors would like to thank X.~Buffat, R.~De Maria, W.~Herr,
B.~Holzer, T.~Pieloni, J. Wenninger (CERN), H.~J.~Kim, N.~Mokhov,
T.~Sen, and V.~Shiltsev (Fermilab) for valuable discussions and
insights. These measurements would not have been possible without the
support of the CERN Operations Group and the Fermilab Accelerator
Division personnel. In particular, we would like to thank S.~Cettour
Cave, A.~Macpherson, D.~Jacquet, M.~Solfaroli Camillocci (CERN),
M.~Convery, C.~Gattuso, and R.~Moore (Fermilab).

Fermilab is operated by Fermi Research Alliance, LLC under Contract
No.~DE-AC02-07CH11359 with the United States Department of Energy.
This work was partially supported by the US LHC Accelerator Research
Program (LARP).

\end{document}